\documentclass[a4paper,11pt]{article}

\usepackage{jheppub} 
\usepackage{slashed}
\usepackage[]{mdframed}
\usepackage[T1]{fontenc} 

\title{\boldmath 
Holographic Krylov complexity in confining gauge theories}
\author[a]{Ali Fatemiabhari,}
\author[b]{Horatiu Nastase,}
\author[c]{Carlos Nunez}
\author[d]{and Dibakar Roychowdhury}

\affiliation[a]{Institute for Theoretical and Mathematical Physics, Lomonosov Moscow State University, 119991 Moscow, Russia}
\affiliation[b]{Instituto de F\'{\i}sica Te\'orica, UNESP-Universidade Estadual Paulista R. Dr. Bento T. Ferraz 271, Bl. II, Sao Paulo 01140-070, SP, Brazil}
\affiliation[c]{Centre for Quantum Fields and Gravity, Department of Physics, Swansea University, Swansea SA2 8PP, United Kingdom}
\affiliation[d]{Department of Physics, Indian Institute of Technology Roorkee,\\Roorkee 247667, Uttarakhand, India}

\abstract{{We study a bulk holographic diagnostic of Krylov (spread) complexity in the Anabal\'on--Ross solitonic background, a top-down Type IIB solution describing a twisted-circle compactification of ${\cal N}=4$ SYM which flows to a gapped, confining, effectively three-dimensional theory in the infrared. Our calculation assumes the recently proposed relation between the rate of spread complexity and the proper radial momentum of an infalling massive probe; it should therefore be read as a controlled bulk test of this proposal, rather than as an independent boundary-field-theory derivation of Krylov complexity. We solve the radial geodesic problem analytically, in the supersymmetric case, in terms of elliptic functions, and verify the result through the UV and IR expansions. The finite end of the space makes the proper momentum finite and leads to bounded, oscillatory behaviour of the associated complexity proxy, in contrast with the unbounded pure-AdS result. We also explain how the oscillation time is controlled by the same geometric scale that sets the Kaluza--Klein and glueball scales. The non-supersymmetric case is treated in the appendix. The result gives a top-down arena in which to sharpen the proposed spread-momentum correspondence and to compare confinement-induced suppression of complexity growth with future direct field-theory calculations.}}
\begin{document} 
\maketitle
\flushbottom
\section{Introduction and General Idea}\label{introgeneral}

Complexity is not a single observable, but rather a family of related diagnostics designed to quantify how difficult it is to prepare, approximate, or time-evolve a state or an operator. In holography, this question has been studied through circuit complexity, path-integral optimization, tensor-network ideas, operator growth, and the complexity-equals-volume/action proposals. In this paper we focus on Krylov, or spread, complexity. Let us recall the minimum amount of formalism needed to make clear what is being computed.

Given a normalized state evolving as $|\psi(t)\rangle=e^{-iHt}|\psi(0)\rangle$, the Lanczos algorithm applied to the Krylov sequence generated by $H$ gives an orthonormal basis $|K_n\rangle$ in which
\begin{equation}
   |\psi(t)\rangle=\sum_{n\geq0}\varphi_n(t)|K_n\rangle,\qquad
   {\cal C}(t)=\sum_{n\geq0} n\, |\varphi_n(t)|^2 .
   \label{eq:introCKstate}
\end{equation}
For operator growth one applies the same construction to the Liouvillian ${\cal L}=[H,\cdot]$. In both cases the Lanczos coefficients encode the spreading of the wavefunction along the Krylov chain, while $C_K(t)$ measures the average Krylov-site number reached at time $t$. The works \cite{Balasubramanian:2022tpr,Caputa:2024sux, Balasubramanian:2023kwd, Caputa:2021sib,Fan:2024iop,Fan:2023ohh,He:2024pox} developed this viewpoint and suggested that Krylov complexity can be used as a geometric and dynamical probe of many-body evolution. A related line of work, involving double-scaled SYK, JT gravity and de Sitter space, includes \cite{Rabinovici:2023yex,Heller:2024ldz,Heller:2025ddj,Fu:2025kkh}; for recent reviews see \cite{Nandy:2024evd,Baiguera:2025dkc,Rabinovici:2025otw}.

The specific holographic input used in this work is the proposal of \cite{Caputa:2024sux}: for suitable heavy states or operators in holographic CFTs, the rate of spread complexity is proportional to the proper radial momentum of an infalling bulk probe,
\begin{equation}
   \frac{d {\cal C}(t)}{dt}=\eta\, P_{\rm proper}(t) ,
   \label{eq:introMomentumProposal}
\end{equation}
where $\eta$ is a normalization fixed by the microscopic definition of the boundary problem. In pure AdS this reproduces the expected conformal behaviour. Our aim is not to derive \eqref{eq:introMomentumProposal} from the boundary theory. Rather, we take it as a well-defined bulk prescription and ask what it predicts in a top-down geometry dual to a confining field theory. This qualification is important: throughout the paper, ``complexity'' denotes the bulk quantity obtained from the proposed spread-momentum map, and a direct boundary Krylov computation remains an important future test.

A natural extension of \cite{Caputa:2024sux} is to replace pure AdS (motivated by Maldacena's conjecture \cite{Maldacena:1997re}) by a confining holographic background. Confinement introduces a mass gap, a discrete low-energy spectrum and a physical infrared scale. These features are expected to suppress the indefinite growth characteristic of conformal dynamics. Related expectations appear in studies of entanglement entropy as a confinement diagnostic \cite{Klebanov:2007ws,Kol:2014nqa, Jokela:2020wgs, Jokela:2025cyz}, in complexity-equals-volume calculations for confining geometries \cite{Fatemiabhari:2024aua,Yang:2023qxx,Frey:2023qdv,Chatzis:2024top,Chatzis:2024kdu,Chatzis:2025dnu}, and in recent Krylov-complexity analyses of confining or gapped systems \cite{Anegawa:2024wov,Jiang:2025wpj}. In particular, \cite{Jiang:2025wpj} emphasizes not merely oscillations, but the suppression of complexity growth relative to an unconstrained, non-confined system. This is the comparison we have in mind below: in the present bulk calculation the confining end of space replaces the unbounded AdS fall by a bounded motion, and hence by a bounded and oscillatory complexity proxy. The oscillating character of the complexity is a direct consequence that when the particle falls-down, complexity increases, whilst when it climbs-up, complexity decreases \cite{Ageev:2018msv, Susskind:2019ddc}.

The top-down background we use is the Anabal\'on--Ross soliton \cite{Anabalon:2021tua}. It is a Type IIB solution describing ${\cal N}=4$ SYM compactified on a spatial circle with an $SO(6)_R$ twist. The ultraviolet is the familiar four-dimensional CFT, while at energies below the inverse circle size the theory flows to an effectively $(2+1)$-dimensional gapped theory. The holographic radial direction ends smoothly at a finite value $z=z_*$, where the $(z,\phi)$ subspace caps off as a cigar. This is the same broad geometric mechanism behind Witten-type confining backgrounds \cite{Witten:1998zw}, but the Anabal\'on--Ross solution has two advantages for our purposes: it is asymptotically AdS$_5$ and therefore directly comparable with the conformal setup of \cite{Caputa:2024sux}, and its metric functions are sufficiently simple that the infalling-particle problem can be solved analytically. This analytic control is harder to maintain in the Witten--Sakai--Sugimoto construction \cite{Sakai:2004cn,Sakai:2005yt} or other confining backgrounds \cite{Klebanov:2000hb, Maldacena:2000yy}, whose UV is not the same four-dimensional conformal theory as we have here. For studies on these other models, see \cite{Fatemiabhari:2026goj}.
 {This paper complements a 
companion paper \cite{Fatemiabhari:2025cyy} in which we calculated the generic evolution of holographic complexity for a four dimensional CFT, ${\cal N}=4$ SYM, and an another 
one on the holographic complexity of one- and two-dimensional quiver CFTs \cite{Fatemiabhari:2025poq}.}

There are two parameters characterising the geometry: $Q$ and $\mu$. They have clear physical roles. The parameter $Q$ is associated with the R-symmetry Wilson line around the compact circle and fixes, in the supersymmetric branch, the IR endpoint $z_*=\frac{1}{Ql}$. The parameter $\mu$ controls a supersymmetry-breaking deformation; $\mu=0$ preserves four supercharges, while $\mu\neq0$ breaks supersymmetry. We keep the main text focused on the supersymmetric case because it is the cleanest analytic setting and already displays the confinement-induced bounded motion. Appendix \ref{sec:ARQmu} gives the corresponding analytic treatment for generic $\mu\neq0$, including the proper momentum and complexity proxy.

The main outcome is the following. We consider a massive particle that is released (with zero initial velocity) from a  UV cutoff \footnote{We UV-regulate the problem, analogous to give a spread in energy to the initial state. ${\cal H}$ is the Hamiltonian of the system, as we explain below.} $z=\epsilon=1/{\cal H}$. As it  reaches the smooth end of space in a finite time $t_e$, reflects, and returns. The proper radial momentum is finite at the tip and changes sign after the reflection. Consequently, the quantity obtained by integrating \eqref{eq:introMomentumProposal} does not grow indefinitely: it oscillates. We interpret this as a bulk signal of the finite interval in radial energy scales available between the UV cutoff and the IR wall. In the boundary language, this should be understood as a large-$N$, semiclassical proxy for the restricted set of states probed by the heavy excitation. This is  {\it not}  a proof that the exact finite-volume Krylov complexity must develop a sharp non-analyticity. The sharp turning points in the plots are probably artifacts of the classical probe limit and should be smoothed by finite-$N$, finite-width wavepacket or direct boundary effects.

The contents of the paper are organized as follows. Section \ref{AnabalonRosssection} reviews the Anabal\'on--Ross solution, explains the twisted compactification and clarifies why the geometry is dual to a gapped confining theory. Section \ref{geodesicsAR} derives the radial geodesic equation, studies its UV and IR limits and presents the exact solution in terms of Jacobi elliptic functions for $\mu=0$. Section \ref{sec:propP} defines the proper radial coordinate, computes the proper momentum, discusses the resulting complexity proxy and explains the relation between the oscillation period and the IR mass scale. Section \ref{QFT-proposal-section} describes what a direct field-theory calculation would require. Appendix \ref{app:Elliptic} collects the elliptic-integral technology used in the analytic solution, and Appendix \ref{sec:ARQmu} treats the generic non-supersymmetric case.

\section{The background of Anabal\'on and Ross}\label{AnabalonRosssection}

We now review the gravitational background and the field-theory interpretation needed for the calculation. The Anabal\'on--Ross geometry \cite{Anabalon:2021tua} is a Type IIB solution with constant dilaton and self-dual five-form flux. It describes ${\cal N}=4$ SYM on $\mathbb{R}^{1,2}\times S^1$, with the circle parametrized by $\phi$, together with a twist by a $U(1)^3\subset SO(6)_R$ R-symmetry connection. Related compactifications and deformations have been studied in \cite{Anabalon:2022aig,Anabalon:2024che,Chatzis:2024kdu,Chatzis:2024top,Chatzis:2025dnu,Castellani:2024ial,Giliberti:2024eii,Fatemiabhari:2024aua,Kumar:2024pcz,Nunez:2023xgl,Barbosa:2024smw,Nunez:2023nnl,Macpherson:2025pqi,Chatzis:2025hek, Anabalon:2026yxk}. We focus on the original Anabal\'on--Ross solution because it is asymptotically AdS$_5$, has a smooth confining cap, and is simple enough to allow the analytic treatment of the probe dynamics.

In coordinates $[t,x_1,x_2,\phi,z,\theta,\psi,\varphi_1,\varphi_2,\varphi_3]$ the metric is
\begin{eqnarray}
& & ds_{10}^2= \frac{l^2}{z^2}\left[-dt^2+dx_1^2+dx_2^2 + f(z) d\phi^2 +\frac{ dz^2}{f(z)}\right]
+ l^2 d\tilde{\Omega}_5^2,\label{ARmetric}\\
& & d\tilde{\Omega}_5^2= d\theta^2+ \sin^2\theta d\psi^2 + \sin^2\theta \sin^2\psi\left(d\varphi_1-A_1 \right)^2+ \nonumber\\
& &~~~~~~~~~~
\sin^2\theta \cos^2\psi\left(d\varphi_2-A_1 \right)^2 +\cos^2\theta \left(d\varphi_3-A_1 \right)^2.\nonumber\\
& & A_1= A_\phi d\phi, ~~A_\phi=Q\left(z^2-{z_*^2} \right),~~~~f(z)= 1-\mu l^2z^4-\left({Ql~z}\right)^6.\nonumber
\end{eqnarray}

The holographic coordinate is $z$, with the UV boundary at $z=0$. The space ends at $z=z_*$, defined by $f(z_*)=0$, so the allowed range is $0\leq z\leq z_*$. Smoothness at the tip fixes the periodicity of the circle,
\begin{equation}
   L_\phi=\frac{4\pi}{|f'(z_*)|}=\frac{2\pi z_*}{2+(Qlz_*)^6},
   \label{eq:LphiARperiod}
\end{equation}
where we have written the positive period explicitly. With this choice the $(z,\phi)$ subspace is locally flat near $z=z_*$ and there is no conical singularity. The relation $f(z_*)=0$ may also be written as
\begin{equation}
   \mu=\frac{1-Q^6l^6z_*^6}{l^2z_*^4} .
\end{equation}
For $\mu=0$ one has $z_*=\frac{1}{Ql}$ and the solution preserves four supercharges; for $\mu\neq0$ supersymmetry is broken.

Let us stress the field-theory interpretation of these statements. The point $z=z_*$ is not a spacetime point where the QFT lives. Rather, it is the endpoint of the holographic RG flow. In fact, the background in eq.(\ref{ARmetric}) describes a stack of D3 branes that wrap a twisted circle. As we lower the energy, KK modes decouple and an RG flow ensues. At high energies, $z\simeq0$, the theory is the four-dimensional conformal ${\cal N}=4$ SYM compactified on the twisted circle. At energies below the Kaluza--Klein scale set by $L_\phi^{-1}$, the massive KK modes decouple and the long-distance dynamics is effectively $(2+1)$-dimensional; the twist induces a Chern--Simons term and preserves four supercharges in the $\mu=0$ branch \cite{Cassani:2021fyv,Kumar:2024pcz,Castellani:2024ial}. This is the intended meaning of saying that the IR theory is effectively three-dimensional.

The reason this background is confining is also geometric. A fundamental string dual to a Wilson--Maldacena loop \cite{Maldacena:1998im} cannot penetrate beyond the smooth cap at $z=z_*$, and the resulting string configuration gives an area law for sufficiently large boundary separation. This comes as a direct application of the formalism in \cite{Sonnenschein:1999if, Nunez_2010}. Equivalently, the low-energy theory has a mass gap and a discrete glueball spectrum. These properties have been verified directly for this class of backgrounds in \cite{Chatzis:2024kdu,Chatzis:2024top,Chatzis:2025dnu,Fatemiabhari:2024aua,Nunez:2023nnl,Nunez:2023xgl}. Thus the AR geometry realizes the same qualitative end-of-space mechanism familiar from Witten-type confining backgrounds, while retaining an asymptotically (SUSY) AdS$_5$ UV that is particularly well suited to comparison with the spread-momentum proposal of \cite{Caputa:2024sux}.

In what follows we compute the bulk side of the Krylov/spread-complexity proposal in this confining background. In the UV we recover the standard AdS behaviour. In the IR the finite endpoint at $z=z_*$ changes the probe motion qualitatively, and it is this change that produces the bounded oscillatory behaviour discussed below.

\section{Geodesic motion in the Anabal\'on-Ross background}\label{geodesicsAR}
As we advanced above, we consider a particle moving in the background of eq.(\ref{ARmetric}). The geodesic is described in terms of the coordinates $[z(t),\phi(t)]$ depending on time. It is consistent to assume that all other coordinates are fixed. Working in Einstein frame (which in this case is equivalent to string frame as the dilaton vanishes), we calculate the induced line element for this particle,
\begin{equation}
ds_{ind}^2 =\frac{l^2 dt^2}{z^2}\left( -1+ (f(z)+ z^2 A_\phi^2 )\dot{\phi}^2+ \frac{\dot{z}^2}{f(z)}\right).   
\end{equation}
This point particle action is 
\begin{align}
\label{eq:ARlag}
    \mathcal{S}_P=- m\int dt \mathcal{L}_P~,~~\mathcal{L}_P=\sqrt{-\det g_{ind}}=\frac{l}{z}\sqrt{1-z^2 M(z)\dot{\phi}^2-\frac{\dot{z}^2}{f(z)}}.
\end{align}
We defined the function  $M(z)=\frac{f(z)}{z^2}+A_\phi(z)^2$ and in what follows we set $ml=1$.

The Lagrangian in eq.\eqref{eq:ARlag} has two conserved quantities. The first one is the angular momentum associated with translations along the $\phi$-cycle,
\begin{align}
\label{eq:Lphi}
    L_\phi = \frac{z M(z)\dot{\phi}}{\sqrt{1-z^2M(z)\dot{\phi^2}-\frac{\dot{z}^2}{f(z)}}}.
\end{align}
%
%
The second conserved quantity is the Hamiltonian (a consequence of  time translation symmetry in the bulk). It reads
\begin{align}
\label{eq:ARH}
    \mathcal{H}=\frac{1}{z \sqrt{1-z^2 M(z)\dot{\phi}^2-\frac{\dot{z}^2}{f(z)}}}.
\end{align}
For simplicity, let us focus on the situation  $\dot{\phi}(t=0)=0$.
Using eq.\eqref{eq:ARH} and $L_\phi = 0$ we obtain
\begin{align}
\label{eq:ARz}
    \dot{z}(t)=\frac{1}{\mathcal{H}z}\sqrt{f(z)(\mathcal{H}^2 z^2-1)}.
\end{align}
%
%
Upon integrating eq.\eqref{eq:ARz}, we find
\begin{eqnarray}
& &     \int dz \frac{z}{\sqrt{-1+\bar\alpha z^2 + \beta z^4 +\gamma z^6 -\zeta z^8}}=\frac{1}{\mathcal{H}}(t-t_0),\label{integral-eq}\\
& &    \bar \alpha = \mathcal{H}^2 ~;~ \beta  = \mu l^2 ~;~ \gamma = Q^6 l^6 - \mu \mathcal{H}^2 l^2~;~ \zeta = Q^6 l^6 \mathcal{H}^2.\nonumber
\end{eqnarray}
It can be checked that the first order equation (\ref{eq:ARz}) and its integral (\ref{integral-eq}) do solve the Euler-Lagrange equations derived from the Lagrangian in eq.(\ref{eq:ARlag}) for the case $\dot{\phi}=0$. Consider the Hamiltonian in eq.(\ref{eq:ARH}),
\begin{align}
    \mathcal{H}=\frac{1}{z \sqrt{1-\frac{\dot{z}^2}{f(z)}}}.
\end{align}
Setting the constant $t_0=0$, we impose the initial conditions  $\dot{z}(t=0)=0$ and $z(t=0)=\epsilon$. Here $\epsilon$ is a very small number, acting as UV cutoff. These initial conditions imply $\mathcal{H}=\frac{1}{\epsilon}$. Hence the Hamiltonian should be large to allow  the initial position of the particle being close to the boundary ($\epsilon\to 0$). In other words, we start with an excitation (an operator, dual to the massive particle) defined in the UV with large energy ${\cal H}$ (large conformal dimension in the UV). This excitation evolves in the QFT. In the dual description, the particle falls in the $z$-direction following the equation (\ref{integral-eq}). First, let us gain some intuition for the motion $z(t)$ using  the UV and IR expansions of  eq.(\ref{integral-eq}). After that, we present an exact analytic solution that confirms the expansions discussed below.

\subsection{Asymptotic expansions  in the SUSY case $(\mu=0)$}\label{sec:ARQAssym}
%
It is instructive to consider the UV and the IR behaviour of the equation for the trajectory (\ref{integral-eq}), using perturbative expansions. This analysis is useful to get an intuitive understanding. After this, we present an exact integration of eq.(\ref{integral-eq}). We focus here on the SUSY case $\mu=0$, which makes the parameters $ \bar \alpha = \mathcal{H}^2 ~;~ \beta  = 0;~ \gamma = Q^6 l^6 ;~ \zeta = Q^6 l^6 \mathcal{H}^2$.
First, consider the expansion of the integrand near $z=\epsilon=1/\mathcal{H}$,
\begin{eqnarray}
& &\frac{z}{\sqrt{-1+\mathcal{H}^2 z^2 +(Ql)^6 z^6 -(Ql)^6\mathcal{H}^2 z^8}}\Bigg|_{z\sim  \frac{1}{\cal H}} =
   \frac{\mathcal{H}^{3/2}}{\sqrt{2} \sqrt{\mathcal{H}^6-(Ql)^6} \sqrt{z-\frac{1}{\mathcal{H}}}}+O(z-\frac{1}{\mathcal{H}})^{1/2},\label{UV-exp-mu=0}\nonumber\\
   \end{eqnarray}
   \begin{eqnarray}
 \frac{1}{\mathcal{H}}(t-t_0) &  & = \int^z d\tilde{z} \frac{\mathcal{H}^{3/2}}{\sqrt{2} \sqrt{\mathcal{H}^6-(Ql)^6} \sqrt{\tilde{z}-\frac{1}{\mathcal{H}}}}+O(z-\frac{1}{\mathcal{H}})^{3/2}\\
 &  &= \frac{\mathcal{H}^{3/2} \sqrt{2 }\sqrt{ z-1/\mathcal{H}}}{\sqrt{\mathcal{H}^6-(Ql)^6}}+O(z-\frac{1}{\mathcal{H}})^{3/2}.\nonumber
\end{eqnarray}
Solving for $z(t)$ with the choice $t_0=0$ we have
\begin{align} \label{eq:z0}
    z(t)= \frac{1}{\mathcal{H}} + \frac{\mathcal{H}^6-(Ql)^6}{2\mathcal{H}^5}t^2 + \cdots
\end{align}
For $\mathcal{H}=\frac{1}{\epsilon}$ and $Q=0$ we have,
\begin{align}
\label{eq:zAR}
    z(t)= \epsilon + \frac{1}{2\epsilon}t^2 + \cdots \sim \sqrt{t^2+\epsilon^2},
\end{align}
 which  matches with the result in equation (11) of the paper  \cite{Caputa:2024sux}.
 
 More revealing  is the expansion close to the end of the space $z=z_*$. In fact,
expanding the integrand close to $z=z_*=\frac{1}{Ql}$, we find
\begin{align}
\frac{z}{\sqrt{-1+\mathcal{H}^2 z^2 +(Ql)^6 z^6 -(Ql)^6\mathcal{H}^2 z^8}}\Bigg|_{z\sim \frac{1}{Ql}} =
   \frac{1}{\sqrt{6} \sqrt{(Ql)^3-\mathcal{H}^2 Ql} \sqrt{z-\frac{1}{Ql}}}+O(z-\frac{1}{Ql})^{1/2}\label{IR-expansion-AR-mu=0}\nonumber\\ 
\end{align}
   \begin{eqnarray}
&  & \frac{1}{\mathcal{H}}(t_e-t) = \int^z d\tilde{z} \frac{1}{\sqrt{6} \sqrt{(Ql)^3-\mathcal{H}^2 Ql} \sqrt{\tilde{z}-\frac{1}{Ql}}}+O(z-\frac{1}{Q})^{3/2} \\
 &  &
=\frac{\sqrt{\frac{2}{3}} \sqrt{z-\frac{1}{Ql }}}{ \sqrt{(Ql)^3-\mathcal{H}^2Ql}}+O(z-\frac{1}{Q})^{3/2}.\nonumber
\end{eqnarray}
The time to reach the end of the space is denoted by $t_e$. Solving for $z(t)$ with $z(t_e)=\frac{1}{Q l}$ we have
\begin{eqnarray}  \label{eq:zte}
    & & z(t)= \frac{1}{Q l}-\frac{3 Ql}{2}\left(1-\frac{ (Ql)^2 }{ \mathcal{H}^2}\right)(t-t_e)^2 + \cdots
\end{eqnarray}
As expected, this result is heavily dependent on the parameter $Q$ and the consequent existence of the end of space at $z_*$.
\\
Having understood  the asymptotic behaviour  of $z(t)$, we study now an exact analytic expression for $z(t)$ solving eqs.(\ref{eq:ARz})-(\ref{integral-eq}).

\subsection{Exact solution in the SUSY case $(\mu=0)$} \label{sec:ARQ}
As above, we consider the case $\mu=0$. The function $f(z)$ is given by $f(z)=1- Q^6 l^6 z^6$. As above we set $\bar \alpha=\mathcal{H}^2$, $\beta = 0$, $\gamma= (Ql)^6$, $\zeta = (Ql)^6\mathcal{H}^2$.
We exactly  perform the integral in eq.\eqref{eq:ARz}, 
\begin{align}
    &\frac{t}{\mathcal{H}} = \int_{1/\mathcal{H}}^z  d \tilde{z} \frac{\tilde{z}}{\sqrt{-1+\mathcal{H}^2 \tilde{z}^2 +(Ql)^6 \tilde{z}^6 -(Ql)^6\mathcal{H}^2 \tilde{z}^8}}\\
    & =\int_{1/\mathcal{H}}^z d\tilde{z} \frac{\tilde{z}}{\sqrt{(\mathcal{H}^2 \tilde{z}^2-1) \left(1-(Ql)^6 \tilde{z}^6\right)}}.
\end{align}
We do a change of variable $u=(Ql)^2 \tilde{z}^2$ and introduce the parameter $\alpha\equiv\mathcal{H}^2/(Ql)^2$.  Now,
\begin{align}
\label{eq:u}
    &\frac{t}{\mathcal{H}} = \frac{1}{2 (Ql)^2\sqrt{\alpha}}\int_{1/\alpha}^{(Ql)^2z^2}  du \frac{1}{\sqrt{(u-1/\alpha)(1-u)(u^2+u+1)}}.
\end{align}

Eq.\eqref{eq:u} is in the form of an incomplete elliptic integral of the first kind as explained in Appendix \ref{app:Elliptic}, in particular in eq.\eqref{eq:elliptic1}. We  use the terminology and abbreviations for the elliptic functions and integrals defined in Appendix \ref{app:Elliptic}, specifically in section \ref{app:exp}, where we carefully evaluate the integral in eq.(\ref{eq:u}) and invert to find $z(t)$. The result for  $z(t)$ is

\begin{align} \label{eq:zfin}
    & z(t)= \frac{1}{Ql} \sqrt{\frac{\frac{A}{\alpha} \left(1+\operatorname{\bf{cn}}(\frac{2 Ql}{g} t,k) \right)+B\left(1-\operatorname{\bf{cn}}(\frac{2 Ql}{g} t,k) \right)}{A\left(1+\operatorname{\bf{cn}}(\frac{2 Ql}{g} t,k)) \right)+B(1-\operatorname{\bf{cn}}\left(\frac{2 Ql}{g} t,k) \right)}}.
\end{align}
We have defined the following constants,
\begin{eqnarray}
   & & g=\left[\frac{\alpha^2}{3\alpha^2+3\alpha+3}   \right]^{\frac{1}{4}},~~~  k^2=\frac{-3(1+\alpha) +2\sqrt{3(\alpha^2+\alpha+1)}}{4\sqrt{3(\alpha^2+\alpha+1)}}. \\
 & & A=\sqrt{3}, \quad B=\sqrt{\frac{\alpha^2+\alpha+1}{\alpha^2}}.\nonumber
\end{eqnarray}
We also denoted by $\operatorname{\bf{cn}}\left(x,k\right)$ the Elliptic cosine function. For definitions, important relations with other Elliptic functions and integrals, see Appendix \ref{app:Elliptic}.
\\
In Figure \ref{fig:ztQ} we plot $z(t)$ for the values $Ql=[\frac{1}{10}, 1, 2]$ and $\mathcal{H}=10$. The results can be compared  with the case of pure AdS in the lower panel of Figure \ref{fig:ztQ}. Note that as $Ql$ decreases, we recover the pure AdS case.

\begin{figure}
    \centering
    \includegraphics[width=0.7\linewidth]{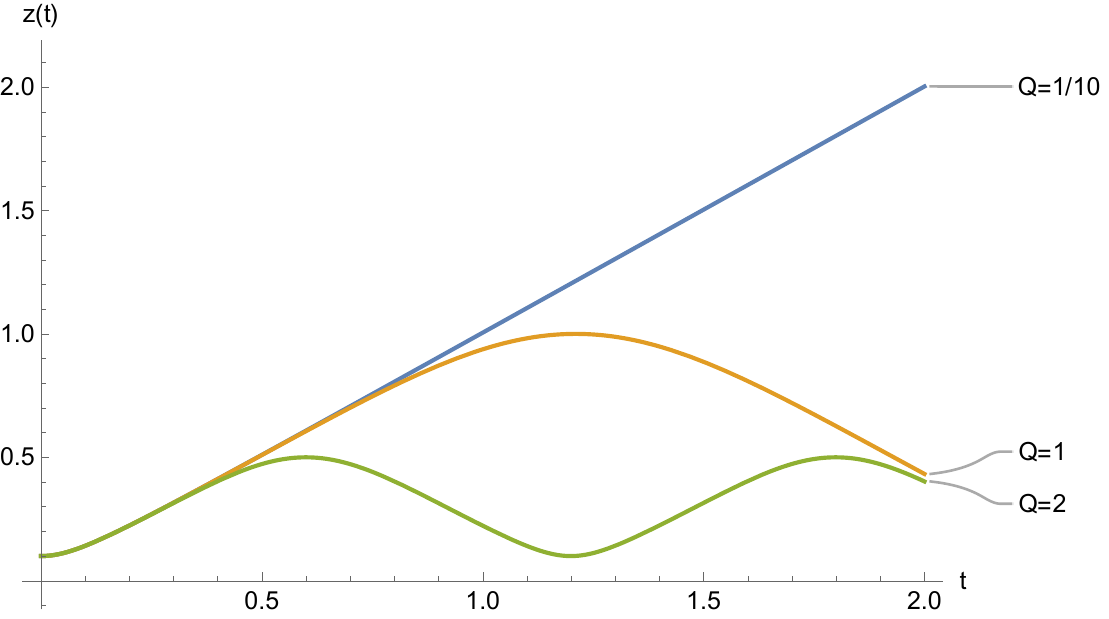}
    
    \vspace{1 cm}
    \includegraphics[width=0.7\linewidth]{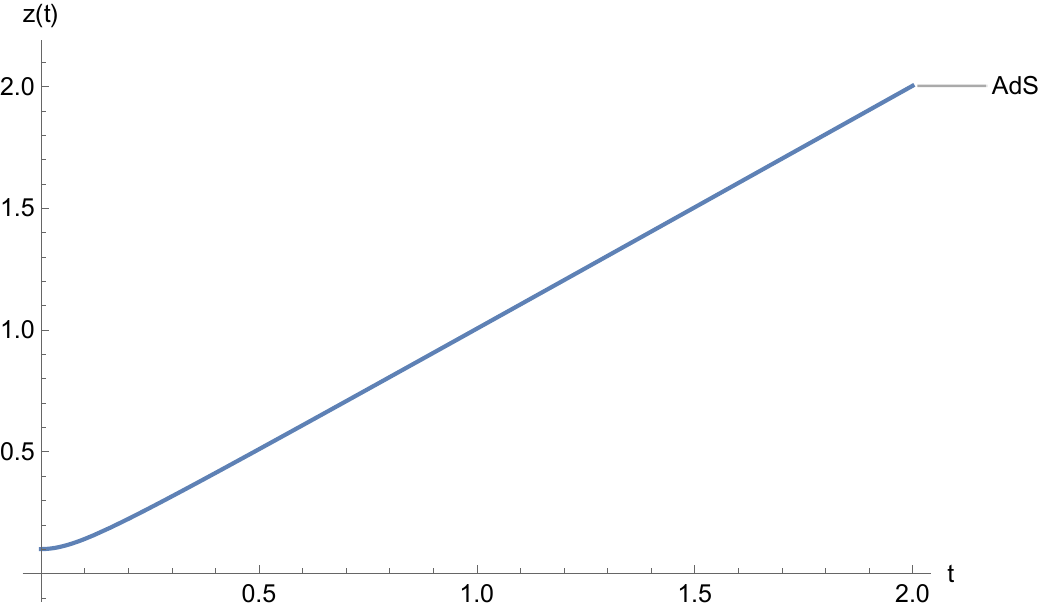}
    \caption{ The function $z(t)$ for choices of $(Ql)=1/10, 1, 2$ and $\mathcal{H}=10$ (Upper panel) and the function $z(t)$ for pure AdS background ($Q=0$)  (Lower panel).}
    \label{fig:ztQ}
\end{figure}

With this analytic solution, we can calculate the time taken to reach the end of the space, imposing that $z(t_e)=\frac{1}{Ql}$. The result is 
\begin{eqnarray}
  & & t_e= \left(\frac{g }{Ql }\right)\!  K\!\left(k\!\right)=
\frac{\sqrt{\alpha}}{Q l [3(\alpha^2+\alpha+1)]^{\frac{1}{4}}}
 K\Bigg[ \frac{\sqrt{-3(1+\alpha)+ 2 \sqrt{3(\alpha^2+\alpha+1)}}}{2[3(\alpha^2+\alpha+1)]^{\frac{1}{4}}}\Bigg].  
\end{eqnarray}
where $K(x)=\int_0^{\frac{\pi}{2}} \frac{dt}{\sqrt{1-x^2 \sin^2 t}}$ is the elliptic integral of the first kind.
We can expand $z(t)$ close to the boundary and the end of space, $z(0)= \frac{1}{\mathcal{H}}$ and $z(t_e)=\frac{1}{Ql}$. We obtain precisely the asymptotic expansions in eqs.(\ref{eq:z0}) and (\ref{eq:zte}) respectively. Furthermore two consistency checks of our solution in eq.(\ref{eq:zfin}) are the following: for $\alpha=1$ (or $\mathcal{H}=Ql$), when the UV-cutoff and the IR-end-of-space do coincide,  we find $z=\frac{1}{Ql}=\epsilon$ and $t_e=0$. On the other hand for $\alpha\to\infty$ (keeping the UV cutoff large and sending $Ql\to 0$), we find that $t_e\sim \frac{1}{Ql}\to\infty$, in agreement for the time taken by a massive particle in AdS to reach the horizon from the boundary.

Considering an expansion of \eqref{eq:zfin} in the limit $Q l \rightarrow 0$ and $\alpha \rightarrow \infty$, one finds
\begin{align}
    z(t)=\frac{1}{\sqrt{\alpha}}\frac{1}{Ql}+\frac{\sqrt{\alpha}Q l}{2}t^2+\cdots
\end{align}
which boils down into an expansion as in pure AdS \eqref{eq:zAR} by remembering the fact that $\alpha = \frac{1}{\epsilon^2 Q^2 l^2}$. In other words, this is the limit in which the cigar transits into a pure AdS and the proper momentum increases indefinitely. All this analysis is in agreement with Figure \ref{fig:ztQ}.

Let us now define the momentum that allows us to calculate the complexity.

\section{The proper momentum and complexity} \label{sec:propP}
We now translate the geodesic solution into the bulk quantity entering the proposed spread-momentum relation \cite{Caputa:2024sux, Fan:2024iop, He:2024pox, Li:2025fqz,Affleck:1991tk,Fatemiabhari:2025cyy}. The relevant momentum is not the canonical momentum conjugate to the coordinate $z$, because $z$ is not a proper radial distance. Instead we define a radial coordinate $\bar\rho$ by requiring that, on a fixed time slice and at fixed boundary and internal coordinates, the line element be $ds^2=d\bar\rho^2$. From \eqref{ARmetric} this gives
\begin{equation}
 ds_{proper}^2= l^2\frac{dz^2}{z^2f(z)}\equiv d\bar{\rho}^2, \quad\Rightarrow \quad -d\bar{\rho}= \frac{l dz}{z\sqrt{f(z)}}. \label{proper-coordinate-def}   
\end{equation}
The sign has been chosen so that $\bar\rho$ increases toward the UV boundary. In the supersymmetric case $f(z)=1-(Qlz)^6$ one finds
\begin{equation}
    z=\frac{1}{Ql \left[ \cosh\left( \frac{3\bar{\rho}}{l}\right)\right]^{\frac{1}{3}}}.\label{zderhobar}
\end{equation}
Thus $\bar\rho\to\infty$ corresponds to the AdS boundary $z\to0$, while $\bar\rho=0$ corresponds to the smooth cap $z=z_*=\frac{1}{Ql}$. In the UV, $z\sim e^{-\frac{\bar\rho}{l}}$, as in the pure-AdS coordinate used in \cite{Caputa:2024sux}. The difference is that in the confining geometry the proper radial coordinate has a lower endpoint.

The conjectural holographic relation is
\begin{align}
    &\partial_t \mathcal{C}(t)   \propto P_{\bar \rho} =P_{z} \frac{\partial \dot z}{\partial \dot {\bar \rho}}=- \frac{\dot{z}}{l \sqrt{f(z)}}\frac{1}{\sqrt{1 -\frac{\dot{z}^2}{f(z)}}} \;.
    \label{eq:PbarRhoDefinition}
\end{align}
Here $\mathcal{C}(t)$ denotes the complexity  obtained from the bulk prescription. The proportionality constant is not fixed by the classical geodesic calculation alone; in a direct boundary calculation it would be fixed by the normalization of the Krylov basis and by the choice of reference state or operator.

Using the asymptotic solution for $z(t)$ we obtain
\begin{align} \label{eq:pbarrhoexp}
 \partial_t \mathcal{C}(t)   \propto &P_{\bar \rho}\Bigg|_{t\sim0} =- \frac{\sqrt{\mathcal{H}^6-(Ql)^6}}{\mathcal{H}^2 l} t +O(t)^{2}\;. \\
 &  P_{\bar\rho}\Bigg|_{t\sim t_e} = \pm\frac{\sqrt{\mathcal{H}^2-(Ql)^2}}{Ql^2} +O(t-t_e)^{2}\;. 
\end{align}
The UV limit agrees with the pure-AdS result of \cite{Caputa:2024sux} when $Ql\to0$. The IR behaviour is different: the proper momentum remains finite at the cap, and the sign changes after reflection. In the same limit $Ql\to0$ the endpoint is pushed to $z_*\to\infty$ and the pure-AdS divergence is recovered.

\begin{figure}
    \centering
\includegraphics[width=0.7\linewidth]{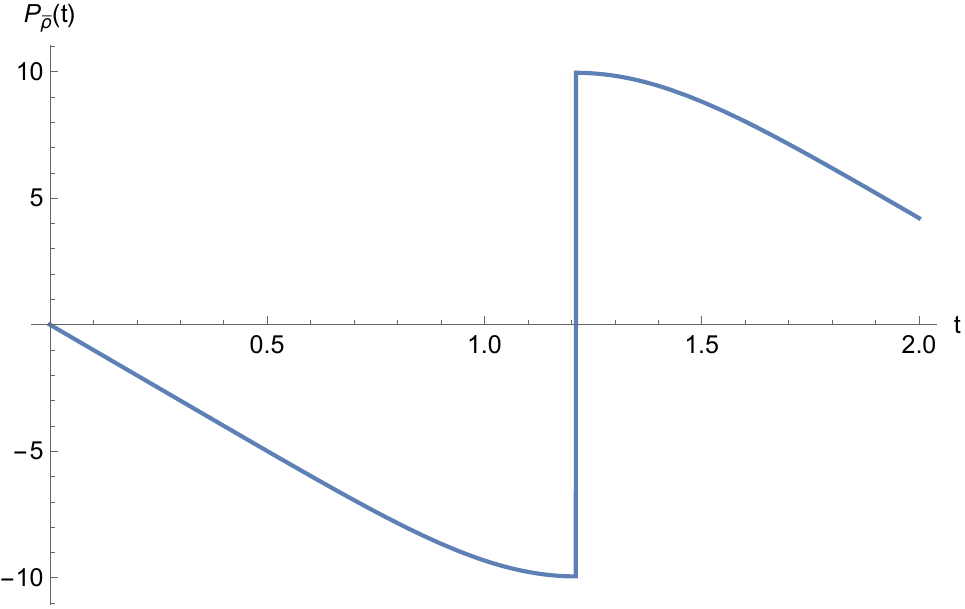}
    \caption{$P_{\bar \rho} (t)$ for $Ql=1$ and $\mathcal{H}=10$. The particle reaches to the end of space at $t_e\simeq 1.21$.}
    \label{fig:PtQ2}
\end{figure}

\begin{figure}
    \centering
    \includegraphics[width=0.7\linewidth]{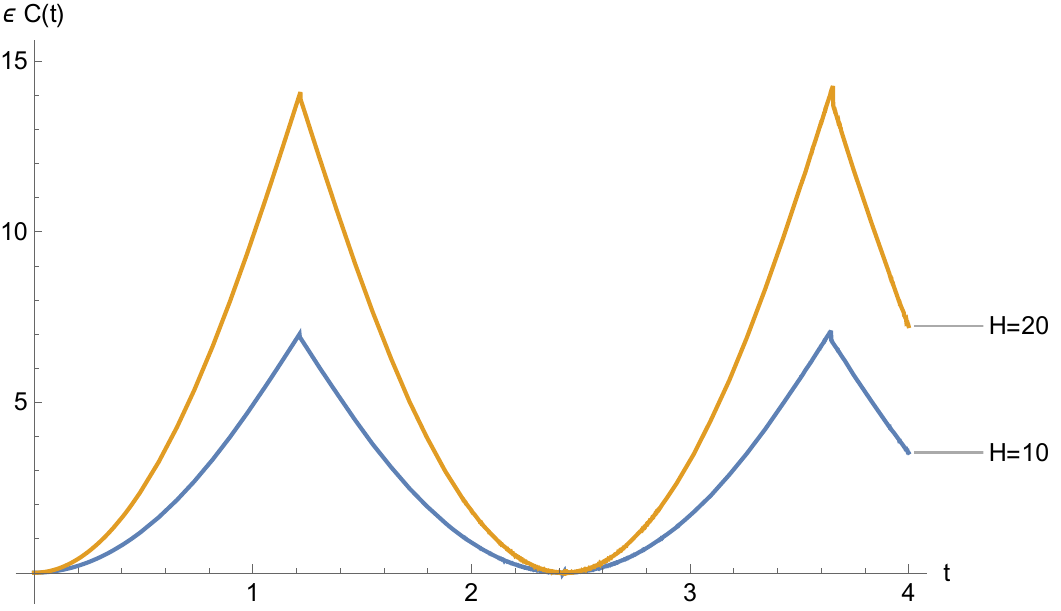}
    \caption{ $\epsilon \;\mathcal{C}(t)$, for $Ql=1$ and $\mathcal{H}=10,20$.}
    \label{fig:CtQ}
\end{figure}

Figure \ref{fig:PtQ2} provides a plot for momentum with choices of $Ql=1$ and $\mathcal{H}=10$.
The complexity  is obtained by integrating the proper momentum. Following the normalization convention used in \cite{Caputa:2024sux}, we write
\begin{equation}
   \mathcal{C}(t)  =- \int dt \frac{P_{\bar \rho}(t)}{\epsilon},
   \label{eq:complexityFromPbarRho}
\end{equation}
with $\epsilon=1/{\cal H}$ in the comparison with the conformal result. Other normalizations are possible until a direct boundary calculation fixes the overall coefficient. Figure \ref{fig:CtQ} displays $\epsilon\,\mathcal{C}(t)$ for two choices of ${\cal H}$.

The physical interpretation in the bulk, is simple. A particle released from the cutoff surface $z=\epsilon$ reaches the smooth cap in a finite time $t_e$. Since the cap is the endpoint of the classical geometry, the particle reflects and returns. The proper momentum therefore changes sign, and the integrated quantity \eqref{eq:complexityFromPbarRho} is bounded and oscillatory rather than monotonically growing. This behaviour should not be read as the statement that oscillations alone are an order parameter for confinement. The more precise statement is that, within the spread-momentum prescription, the confining geometry suppresses the unbounded growth found in pure AdS (where ${\cal C}(t)\sim t^2$) and replaces it by a finite-amplitude oscillation. This is the point of contact with \cite{Jiang:2025wpj}, where the confining deformation reduces complexity growth relative to the non-confined model, and with \cite{Anegawa:2024wov}, where Krylov complexity is used to diagnose confinement-deconfinement physics.

The sharpness of the turning point deserves a comment. In the classical bulk approximation the probe is pointlike and the end of space is a perfectly reflecting wall; this produces a non-analytic change of slope in the momentum and in the integrated complexity. A finite-volume or finite-dimensional quantum system need not show such a sharp cusp. We expect the exact boundary answer, at finite $N$ or for a wavepacket of finite radial width, to smooth the non-analyticity into a rapid crossover or a sequence of avoided recurrences. The non-analyticity in the figures is therefore best understood as a large-$N$, classical-geodesic idealization of a bounded Krylov evolution.

It is also useful to identify the scale controlling the oscillations. In the supersymmetric branch $z_*=\frac{1}{Ql}$ and
\begin{equation}
   L_\phi=\frac{2\pi z_*}{3},\qquad
   M_{\rm KK}\sim L_\phi^{-1}=\frac{3}{2\pi z_*} .
   \label{eq:MKKscale}
\end{equation}
The glueball masses in such capped geometries are expected to scale as $M_n=c_n/z_*$, where the dimensionless constants $c_n$ follow from the appropriate fluctuation problem. The exact result \eqref{eq:zfin} gives
\begin{equation}
   t_e=\frac{g}{Ql}K(k)= z_*\, g\, K(k),
   \label{eq:teScaleRewritten}
\end{equation}
with $g$ and $k$ depending only on the dimensionless combination ${\cal H}z_*$. Therefore the characteristic oscillation time, with period $T_{\rm osc}\simeq2t_e$ for the complete return to the cutoff surface, is controlled by the same IR scale that sets $M_{\rm KK}$ and the glueball spectrum. A quantitative test of a relation such as $T_{\rm osc}^{-1}\sim M_{\rm gap}$ requires the direct computation of the relevant glueball masses in this precise background. In  Section \ref{QFT-proposal-section},  we explain what such a comparison would involve.

Finally, increasing ${\cal H}$ moves the initial cutoff surface closer to the UV boundary and increases the maximum value reached by the complexity, as shown in Figure \ref{fig:CtQ}. In the boundary interpretation this corresponds to allowing the initial excitation to explore a larger effective Krylov subspace before it feels the IR cap. The same qualitative behaviour persists in the generic $\mu\neq0$ solution discussed in Appendix \ref{sec:ARQmu}, although the detailed time scale and amplitude depend on the supersymmetry-breaking parameter.

\subsection{A field theoretical approach? }\label{QFT-proposal-section}
A precise boundary (QFT) calculation of the Krylov spread complexity for our system is beyond the capabilities of this work. Let us spell out what a direct boundary calculation would require. At very high energies the theory is ${\cal N}=4$ SYM compactified on a twisted $S^1$. The perturbative Lagrangian and KK spectrum have been described in \cite{Kumar:2024pcz,Castellani:2024ial}. In that regime one could, in principle, compute two-point functions for the KK modes, extract the survival amplitude, construct the Lanczos coefficients, and determine the spread complexity using methods such as those in \cite{Jefferson:2017sdb,Bhattacharyya:2018bbv,Moghimnejad:2021rqe,Avdoshkin:2022xuw,Nandy:2024evd,Anegawa:2024wov}.

The IR theory dual to the capped geometry is more complicated. Its excitations are color-singlet composites: scalar, vector and tensor glueballs, together with their fermionic partners in the supersymmetric branch. A useful toy model, close in spirit to \cite{Anegawa:2024wov}, would replace the confining theory by a tower of weakly coupled modes,
\begin{equation}
L\sim \sum_{n=1}^\infty (\partial \phi_n)^2 - m_n^2\phi_n^2. \label{gluelag} 
\end{equation}
Given the correlators of these modes one could construct the Krylov wavefunctions and compute
\begin{equation}
   {\cal C§}(t)=\sum_{n\geq0} n\,|\psi_n(t)|^2.
\end{equation}
However, for the actual Anabal\'on--Ross compactification the masses $m_n$ and the relevant operator overlaps are not yet known in sufficient detail. The required calculation is a fluctuation analysis of the five-dimensional gauged supergravity whose uplift gives \eqref{ARmetric}: one must expand the metric, gauge fields and scalars to quadratic order, form gauge-invariant combinations, diagonalize the coupled radial equations and impose regularity at the cap and normalizability at the boundary. Similar computations for related systems were carried out in \cite{Nunez:2023nnl,Nunez:2023xgl,Fatemiabhari:2024aua,Fatemiabhari:2024lct}.

For this reason the present paper does not claim a boundary derivation of the Krylov complexity of the compactified gauge theory. The result is a bulk prediction, motivated by the spread-momentum proposal, which is qualitatively consistent with the suppression and oscillatory behaviour observed in confining or gapped models \cite{Anegawa:2024wov,Jiang:2025wpj}. A direct boundary computation of the glueball correlators and Lanczos data would be the cleanest next test of the proposal.
 Let us close this paper with some summary and conclusions.

\section{Conclusions and Outlook}\label{conclsect}
We have studied a bulk holographic proxy for Krylov/spread complexity in a top-down confining background. The gravitational model is the Anabal\'on--Ross soliton, dual to a twisted compactification of ${\cal N}=4$ SYM which is conformal in the UV and flows to a gapped, effectively three-dimensional theory in the IR. The calculation uses the proposal of \cite{Caputa:2024sux} that the time derivative of spread complexity is proportional to the proper radial momentum of an infalling massive probe, see \cite{Fan:2024iop, He:2024pox, Li:2025fqz,Fatemiabhari:2025cyy} for further elaborations on this proposal. Thus the main result of this work is a controlled bulk prediction, not a direct field-theory computation of Krylov complexity.

The main technical achievement is the analytic solution of the radial geodesic problem. In the supersymmetric branch $\mu=0$ the trajectory $z(t)$ can be written exactly in terms of Jacobi elliptic functions. The UV expansion reproduces the pure-AdS result, while the IR expansion is controlled by the smooth endpoint $z=z_*$. The corresponding proper momentum is finite at the cap and changes sign when the probe reflects. Consequently, the  complexity  is bounded and oscillatory.

This result should be interpreted with some care. We do not claim that the mere presence of oscillations is, by itself, a universal order parameter for confinement. Rather, the robust statement is comparative: relative to the conformal AdS calculation, the confining end of space suppresses the indefinite growth of the proper momentum and replaces it by a finite-amplitude recurrence. This is qualitatively consistent with recent field-theory and lattice studies in which confinement or a mass gap reduces the growth of Krylov complexity. In our setup the characteristic time is $t_e=z_* gK(k)$, so the oscillation period is controlled by the same IR scale that sets the inverse KK scale and the expected glueball masses. A precise numerical relation between the oscillation period and the lightest glueball mass requires the fluctuation-spectrum calculation described in Section \ref{QFT-proposal-section}.

The non-analyticity at the turning point is also a feature of the classical bulk approximation. The point particle reflects from a perfectly smooth geometry, and this idealization produces a sharp (instantaneous) change of sign in the momentum. In a finite-$N$ boundary theory, or for a bulk wavepacket of finite width, we expect this behaviour to be smoothed-out. The non-analyticity should therefore be viewed as the semiclassical limit of a bounded recurrence, not as a literal prediction of a cusp in every microscopic definition of Krylov complexity.

Appendix \ref{sec:ARQmu} extends the analytic treatment to generic $\mu\neq0$, where supersymmetry is broken. We kept the supersymmetric case in the main text because it is the cleanest setting in which to present the exact solution and the physical mechanism. The appendix shows that the same method applies more generally, and it provides the starting point for a systematic comparison of supersymmetric and non-supersymmetric flows.

Several directions are natural. First, one should compute the glueball spectrum and the relevant two-point functions of the Anabal\'on--Ross compactification, construct the corresponding Lanczos coefficients and compare the resulting boundary Krylov complexity with the bulk prediction. Second, it would be valuable to scan the full $(Q,\mu,{\cal H})$ parameter space, using the analytic formulas derived here, to determine how the maximum complexity, recurrence time and proper-momentum profile depend on the IR scale and on supersymmetry breaking. Third, one can include angular momentum along the compact circle or R-charge on the internal space, thereby probing operators with additional quantum numbers. Fourth, string and brane probes should be studied, since extended objects represent classes of boundary operators whose dynamics cannot be captured by point particles. Finally, the same calculation can be repeated in related uplifted solitons and confining compactifications \cite{Anabalon:2024che,Fatemiabhari:2024aua,Chatzis:2024kdu,Chatzis:2024top,Chatzis:2025dnu}, allowing one to distinguish universal confinement effects from model-dependent features of the Anabal\'on--Ross geometry.

We hope that the analytic results presented here will be useful as a benchmark for future field-theory, lattice and holographic studies of Krylov complexity in gapped large-$N$ systems.

\section*{Acknowledgments}

For discussions, for comments on the manuscript, and for sharing their ideas with us, we wish to thank: Dmitry Ageev, Nicol\'o Bragagnolo, Alfonso Ramallo, Ricardo Terrazas. We thank the anonymous referee, whose comments helped us improve the presentations of the contents of this work. 
The work of HN is supported in part by  CNPq grant 304583/2023-5 and FAPESP grant 2019/21281-4.
HN would also like to thank the ICTP-SAIFR for their support through FAPESP grant 2021/14335-0. 
DR would like to acknowledge the Mathematical Research Impact Centric Support (MATRICS) grant (MTR/2023/000005) received from ANRF, India

\vspace{0.5cm}

{\bf Open Access Statement} --- For the purpose of open access, the authors have applied a Creative Commons Attribution (CC BY) license to any Author Accepted Manuscript version arising. 

\vspace{0.5cm}


\vspace{0.5cm}

\hrule

\appendix

\section{Elliptic Integrals} \label{app:Elliptic}
In this appendix, we present the prescription to calculate certain integrals appearing in the body of the paper. They will be used extensively to calculate different observables.

We will introduce terminology and abbreviations for the elliptic functions and integrals which are required to perfom our calculations, following \cite{Byrd:1971bey}. 

For the integral of the form 
\begin{equation}
    \int_b^y \frac{d u}{\sqrt{(a-u)(u-b)(u-c)(u-\bar{c})}}, \label{eq:elliptic1}
\end{equation}
with $a, b$ real, $a \geq y>b, c, \bar{c}$ complex, which is in the form of an incomplete elliptic integral of the first kind. One finds
\begin{align}
&\sqrt{(a-u)(u-b)(u-c)(u-\bar c)}=\sqrt{(a-u)(u-b)\left[\left(u-b_1\right)^2+a_1^2\right]} ; \quad \nonumber\\
&b_1=\frac{c+\bar{c}}{2}, \quad a_1^2=-\frac{(c-\bar{c})^2}{4}. \label{eq:ellipticpara1}
\end{align}
Now one can define a change of variable from $u$ to $\bar u$ with the following relations
\begin{align}
 &\operatorname{\bf{cn}} (\bar u,k)=\frac{(a-u) B-(u-b) A}{(a-u) B+(u-b) A}, \quad A^2=\left(a-b_1\right)^2+a_1^2, \nonumber\\
&B^2=\left(b-b_1\right)^2+a_1^2, \quad g=\frac{1}{\sqrt{A B}}, \quad k^2=\frac{(a-b)^2-(A-B)^2}{4 A B}. \label{eq:ellipticpara2}
\end{align}
Here, `$\operatorname{\bf{cn}}$' is the elliptic cosine function. The upper limit of the integral in eq.\eqref{eq:elliptic1}, $y$, gets mapped to a new parameter that we would like to call it $u_1$. Later, it turns out to be convenient to have $\varphi$ defined below in terms of the $u_1$ variable or originally $y$,
\begin{align}
&\operatorname{\bf{cn}} (u_1,k)=\cos \varphi, \quad \varphi=\operatorname{am} (u_1,k)=\cos ^{-1}\left[\frac{(a-y) B-(y-b) A}{(a-y) B+(y-b) A}\right], \label{eq:ellipticpara3}
\end{align}
where `$\operatorname{am}$', is the Jacobi amplitude defined here as $\operatorname{am} (x,k) \equiv \cos ^{-1}{\operatorname{\bf{cn}}} (x,k) $. 

Then the integral in eq.\eqref{eq:elliptic1} takes the following form, 
\begin{align}
\int_b^y \frac{d u}{\sqrt{(a-u)(u-b)(u-c)(u-\bar{c})}}=g \int_0^{u_1} d \bar u=g u_1 =& g \operatorname{\bf{cn}}^{-1}(\cos \varphi, k) \nonumber\\
 =&g \mathbf{F}(\varphi, k) . \label{eq:elliptic1F}
\end{align}
The definition of `$\operatorname{am}$' and `$\operatorname{\bf{cn}}$' functions are implicitly given among the above relations. Here, `$\mathbf{F}$' is the incomplete elliptic integral of the first kind defined 
below,
\begin{align} \label{eq:defF}
    \int_0^\varphi \frac{d \theta}{\sqrt{1-k^2 \sin^2{\theta}}}\equiv \mathbf{F}(\varphi,k).
\end{align}
Note the appearance of $k^2$ in the denominator of the integral, but $k$ in the argument of the $\mathbf{F}$ function, as different conventions are employed in the literature. Specially, Mathematica software's definition requires $k^2$ as the second argument of the `$\mathbf{F}$' function to match our definition which has only $k$ appearing as the second argument.



\subsection{An explicit calculation} \label{app:exp}


In this section we give a detailed derivation of the result for the $z(t)$ function presented in eq.\eqref{eq:zfin} in the body of the paper.

The integral
\begin{align}
    &\frac{t}{\mathcal{H}} = \frac{1}{2 (Ql)^2\sqrt{\alpha}}\int_{1/\alpha}^{(Ql)^2z^2}  du \frac{1}{\sqrt{(u-1/\alpha)(1-u)(u^2+u+1)}},
\end{align}
also given in eq.\eqref{eq:u}, is in the form of an incomplete elliptic integral of the first kind as explained above.

Now we define the new parameters given in eqs.~(\ref{eq:ellipticpara1}-\ref{eq:ellipticpara3}) required in the final integral result as
\begin{align}
&a=1, \quad b= 1/\alpha, \quad c=e^{2\pi i /3},\quad \bar c=e^{-2\pi i /3}; \quad b_1=-1/2, \quad a_1^2=3/4,  \nonumber\\
& A^2=3, \quad B^2=\frac{\alpha^2+\alpha+1}{\alpha^2}, \quad g=\left[\frac{\alpha^2}{3\alpha^2+3\alpha+3}   \right]^{\frac{1}{4}},\nonumber\\
& k^2=\frac{-3(1+\alpha) +2\sqrt{3(\alpha^2+\alpha+1)}}{4\sqrt{3(\alpha^2+\alpha+1)}}, \nonumber\\
&\varphi=\operatorname{am} (u_1,k)=\cos ^{-1}\left[\frac{(1-(Ql)^2z^2) B-((Ql)^2z^2-1/\alpha) A}{(1-(Ql)^2z^2) B+((Ql)^2z^2-1/\alpha) A}\right] . \label{eq:para}
\end{align}
Here, `$\operatorname{am}$' is the Jacobi amplitude defined in the previous section. 
The variables $\varphi$ and $u_1$ are given implicitly in terms of $z$ in the last relation of the eq.\eqref{eq:para}.

Then, using eq. \eqref{eq:elliptic1F}, one finds  
\begin{align} \label{eq:ellipticF}
\frac{t}{\mathcal{H}} = \frac{1}{2 (Ql)^2\sqrt{\alpha}}\int_{1/\alpha}^{(Ql)^2z^2}  du \frac{1}{\sqrt{(u-1/\alpha)(1-u)(u^2+u+1)}}  &=\frac{g}{2 (Ql)^2\sqrt{\alpha}} u_1 \\
 &  
 =\frac{g}{2 (Ql)^2\sqrt{\alpha}} \mathbf{F}(\varphi, k) .\nonumber
 \end{align}
 We insert $\varphi$ and $k$ from eq.\eqref{eq:para} and obtain
 \begin{align}
 &  \frac{t}{\mathcal{H}}=\frac{g}{2 (Ql)^2\sqrt{\alpha}} \mathbf{F}(\cos ^{-1}\left[\frac{(1-(Ql)^2 z^2) B-((Ql)^2 z^2-1/\alpha) A}{(1-(Ql)^2 z^2) B+((Ql)^2 z^2-1/\alpha) A}\right], \sqrt{\frac{(1-1/\alpha)^2-(A-B)^2}{4 A B}}),
\end{align}
`$\mathbf{F}$' being the incomplete elliptic integral of the first kind defined in eq.\eqref{eq:defF}. Using the relations among elliptic integrals and Jacobi functions one can invert the above relations and solve for $z(t)$. One has

\begin{align}
    \varphi= \operatorname{am}(u_1,k) &\rightarrow \operatorname{\bf{cn}}(u_1,k)\equiv \cos \operatorname{am}(u_1,k)= \cos \varphi =\frac{(1-(Ql)^2 z^2) B-((Ql)^2 z^2-1/\alpha) A}{(1-(Ql)^2 z^2) B+((Ql)^2 z^2-1/\alpha) A}\nonumber \\
    & z(t)= \frac{1}{Ql }\sqrt{\frac{\frac{A}{\alpha}\left(1+\operatorname{\bf{cn}}(u_1,k) \right)+B(1-\operatorname{\bf{cn}}(u_1,k))}{A(1+\operatorname{\bf{cn}}(u_1,k))+B(1-\operatorname{\bf{cn}}(u_1,k))}}.
\end{align}
Recalling $u_1=\frac{2 (Ql)}{g} t$ from eq.\eqref{eq:ellipticF}, we get
\begin{align}
    & z(t)= \frac{1}{(Ql)} \sqrt{\frac{A/\alpha(1+\operatorname{\bf{cn}}(\frac{2 (Ql)}{g} t,k)+B(1-\operatorname{\bf{cn}}(\frac{2 (Ql)}{g} t,k))}{A(1+\operatorname{\bf{cn}}(\frac{2 (Ql)}{g} t,k))+B(1-\operatorname{\bf{cn}}(\frac{2 (Ql)}{g} t,k))}},
\end{align}
which is the desired final result given in the body of the paper.

\section{Exact solution for $\mu\neq0$ and $Q\neq0$} \label{sec:ARQmu}
In this section, we consider the generic case $\mu\neq0,Q\neq0$, where the function $f(z)$ is given by $f(z)=1- \mu l^2 z^4-Q^6 l^6 z^6$. Furthermore, we set $\bar \alpha = \mathcal{H}^2 ~,~ \beta  = \mu l^2 ~,~ \gamma = Q^6 l^6 - \mu \mathcal{H}^2 l^2~,~ \zeta = Q^6 l^6 \mathcal{H}^2.$

Here we perform the integral in eq.\eqref{eq:ARz}, exactly.
We have 
\begin{align}
    &\frac{t}{\mathcal{H}} = \int_{1/\mathcal{H}}^z  d\tilde{z} \frac{\tilde{z}}{\sqrt{f(\tilde{z})(\mathcal{H}^2 \tilde{z}^2-1)}}\\
    & =\int_{1/\mathcal{H}}^z d\tilde{z} \frac{\tilde{z}}{\sqrt{(\mathcal{H}^2 \tilde{z}^2-1) \left(1-\mu l^2 \tilde{z}^4-(Ql)^6 \tilde{z}^6\right)}}.
\end{align}
We do a change of variable $u=(Ql)^2 \tilde{z}^2$ and introduce the parameter $\alpha\equiv\mathcal{H}^2/(Ql)^2$.  Now,
\begin{align}
\label{eq:uQmu}
    &\frac{t}{\mathcal{H}} = \frac{1}{2 (Ql)^2\sqrt{\alpha}}\int_{1/\alpha}^{(Ql)^2z^2}  du \frac{1}{\sqrt{(u-1/\alpha)(-u^3-\frac{\mu}{Q^4l^2} u+1)}}.
\end{align}

Eq.\eqref{eq:uQmu} is in the form of an incomplete elliptic integral of the first kind. Now, we use the same relations as the ones in the previous Appendix \ref{app:Elliptic}. 

We need to find the roots of the cubic function $-u^3-\frac{\mu}{Q^4l^2} u^2+1\equiv(a-u)(u-c)(u-\bar{c})$ appearing in the denominator.
The roots are calculable analytically; here we mention the real root denoted as $a$, related to the position of the end of space, and $b_1$ and $a_1$ related to the sum and difference of complex roots.
\begin{align}
    a &= \frac{\sqrt[3]{-\frac{2 \mu ^3}{Q^{12}l^6}-3 \sqrt{81-\frac{12 \mu ^3}{Q^{12}l^6}}+27}+\sqrt[3]{3 \left(\sqrt{81-\frac{12 \mu ^3}{Q^{12}l^6}}+9\right)-\frac{2 \mu ^3}{Q^{12}l^6}}}{3 \sqrt[3]{2}}-\frac{\mu }{3 Q^4l^2}\nonumber\\
    b_1&=\frac{c+\bar{c}}{2}= -\frac{\sqrt[3]{-\frac{2 \mu ^3}{Q^{12}l^6}-3 \sqrt{81-\frac{12 \mu ^3}{Q^{12}l^6}}+27}+\sqrt[3]{3 \left(\sqrt{81-\frac{12 \mu ^3}{Q^{12}l^6}}+9\right)-\frac{2 \mu ^3}{Q^{12}l^6}}}{6 \sqrt[3]{2}}-\frac{\mu }{3 Q^4l^2}, \nonumber\\
a_1^2& =-\frac{(c-\bar{c})^2}{4}=\frac{\left(\sqrt[3]{-\frac{2 \mu ^3}{Q^{12}l^6}-3 \sqrt{81-\frac{12 \mu ^3}{Q^{12}l^6}}+27}-\sqrt[3]{3 \left(\sqrt{81-\frac{12 \mu ^3}{Q^{12}l^6}}+9\right)-\frac{2 \mu ^3}{Q^{12}l^6}}\right)^2}{12*2^{2/3}}.
\end{align}
These relations reduce to the relevant equations in section \ref{app:exp} in the $\mu\to0$ limit.
Now, one can define the required parameters, which are kept implicit for the peace of mind of the reader.
\begin{align}
  &  \quad A^2=\left(a-b_1\right)^2+a_1^2,\;\; 
B^2=\left(b-b_1\right)^2+a_1^2, \quad g=\frac{1}{\sqrt{A B}}, \quad k^2=\frac{(a-b)^2-(A-B)^2}{4 A B}, \nonumber\\
&b= 1/\alpha, \quad \operatorname{\bf{cn}} u_1=\cos \varphi, \quad \varphi=\operatorname{am} u_1=\cos ^{-1}\left[\frac{(a-(Ql)^2z^2) B-((Ql)^2z^2-b) A}{(a-(Ql)^2z^2) B+((Ql)^2z^2-b) A}\right] .
\end{align}

\begin{figure}
    \centering
    \includegraphics[width=0.7\linewidth]{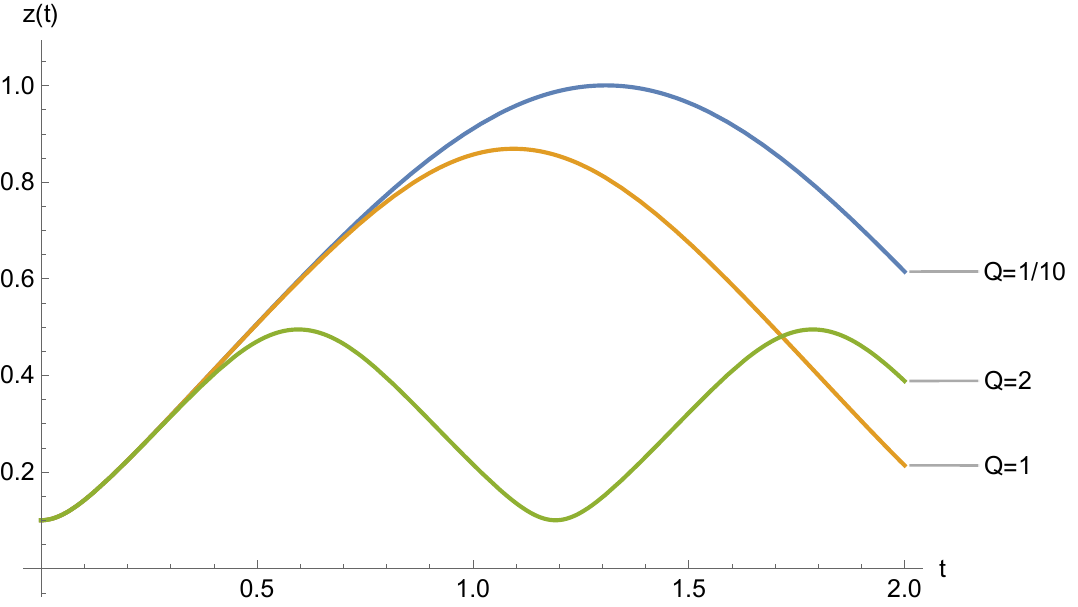}
    \caption{$z(t)$ for various values of Q and $\mu=1.$}
    \label{fig:ztQmu}
\end{figure}

Then, 
\begin{align} \label{eq:ellipticFQmu}
\frac{t}{\mathcal{H}} &= \frac{1}{2 (Ql)^2\sqrt{\alpha}}\int_{1/\alpha}^{(Ql)^2z^2}  du \frac{1}{\sqrt{(u-1/\alpha)(a-u)(u-c)(u-\bar{c})}}  =\frac{g}{2 (Ql)^2\sqrt{\alpha}} u_1  \nonumber\\
 &  =\frac{g}{2 (Ql)^2\sqrt{\alpha}} \operatorname{\bf{cn}}^{-1}(\cos \varphi, k) =\frac{g}{2 (Ql)^2\sqrt{\alpha}} \mathbf{F}(\varphi, k) \nonumber\\
 &  =\frac{g}{2 (Ql)^2\sqrt{\alpha}} \mathbf{F}(\cos ^{-1}\left[\frac{(a-(Ql)^2 z^2) B-((Ql)^2 z^2-1/\alpha) A}{(a-(Ql)^2 z^2) B+((Ql)^2 z^2-1/\alpha) A}\right], \sqrt{\frac{(a-1/\alpha)^2-(A-B)^2}{4 A B}})
\end{align}
Using the relations among elliptic integrals and Jacobi functions, one can invert the above relations and solve for $z(t)$. One has 
\begin{align}
    &\varphi= \operatorname{am}(u_1,k) \rightarrow \operatorname{\bf{cn}}(u_1,k)\equiv \cos \operatorname{am}(u_1,k)= \frac{(a-(Ql)^2 z^2) B-((Ql)^2 z^2-1/\alpha) A}{(a-(Ql)^2 z^2) B+((Ql)^2 z^2-1/\alpha) A}\nonumber \\
    & z(t)= 1/(Ql) \sqrt{\frac{A/\alpha(1+\operatorname{\bf{cn}}(u_1,k))+a B(1-\operatorname{\bf{cn}}(u_1,k))}{A(1+\operatorname{\bf{cn}}(u_1,k))+B(1-\operatorname{\bf{cn}}(u_1,k))}}.
\end{align}
Remembering $u_1=\frac{2 (Ql)}{g} t$ from eq.\eqref{eq:ellipticFQmu}, we get
\begin{align}
    & z(t)= \frac{1}{(Ql)} \sqrt{\frac{A/\alpha(1+\operatorname{\bf{cn}}(\frac{2 (Ql)}{g} t,k))+a B(1-\operatorname{\bf{cn}}(\frac{2 (Ql)}{g} t,k))}{A(1+\operatorname{\bf{cn}}(\frac{2 (Ql)}{g} t,k))+B(1-\operatorname{\bf{cn}}(\frac{2 (Ql)}{g} t,k))}}.
\end{align}
This is the desired result and we plot $z(t)$ for $Q=1/10, 1, 2$ and $l=1, \mu=1,\mathcal{H}=10$ in Figure \ref{fig:ztQmu}. 

We can expand $z(t)$ close to boundary and end of space, $z(0)= 1/\mathcal{H}$ and $z(t_e)=\sqrt{a}/(Ql)$, where 
\begin{eqnarray}
   t_e= \frac{g K\left(k\right)}{(Ql) },
\end{eqnarray}
with $K$ being the elliptic integral of the first kind.

One gets
\begin{align}
    z(t)\Bigg|_{t\sim0}&= \frac{1}{\mathcal{H}} + \frac{(Ql) (a-b) \left(a_1^2+(b-b_1)^2\right)}{2 \sqrt{b}}t^2 + O(t)^3= \frac{1}{\mathcal{H}} + \frac{(Ql) f(\frac{\sqrt{b}}{(Ql)})}{2 \sqrt{b}}t^2+ O(t)^3,\\
    z(t)\Bigg|_{t\sim t_e}&= \frac{\sqrt{a}}{(Ql)}-\frac{(Ql) (a-b) \left(a^2-2 a\;b_1+a_1^2+b_1^2\right)}{2 \sqrt{a}} (t-t_e)^2 + O(t-t_e)^3, \\
\end{align}
matching exactly with eq.~\eqref{eq:z0} and eq.~\eqref{eq:zte} in the $\mu \rightarrow 0$ limit. 


The expansion of the proper momentum in the $\bar \rho$ coordinate,
\begin{equation}
 -d\bar{\rho}= \frac{l dz}{z\sqrt{f(z)}},
\end{equation}
close to the boundary and end of the space is 
\begin{align}
 \partial_t \mathcal{C}(t)   \propto &P_{\bar \rho}\Bigg|_{t\sim0} =-\frac{\sqrt{a-b}\sqrt{a_1^2+(b-b_1)^2}(Ql)}{\sqrt{b}}t +O(t)^{2}=-\frac{\sqrt{f(\sqrt{b}/(Ql))}}{\sqrt{b}/(Ql)}t +O(t)^{2}\;. \\
 &  P_{\bar\rho}\Bigg|_{t\sim t_e} = \pm\frac{\sqrt{a-b}}{\sqrt{b}} +O(t-t_e)^{2}\;. 
\end{align}
Remembering $\sqrt{b}/(Ql)=1/\mathcal{H}$, this matches with the previous results for $\mu\rightarrow0$ as provided in the previous sections.

\begin{figure}
    \centering
    \includegraphics[width=0.7\linewidth]{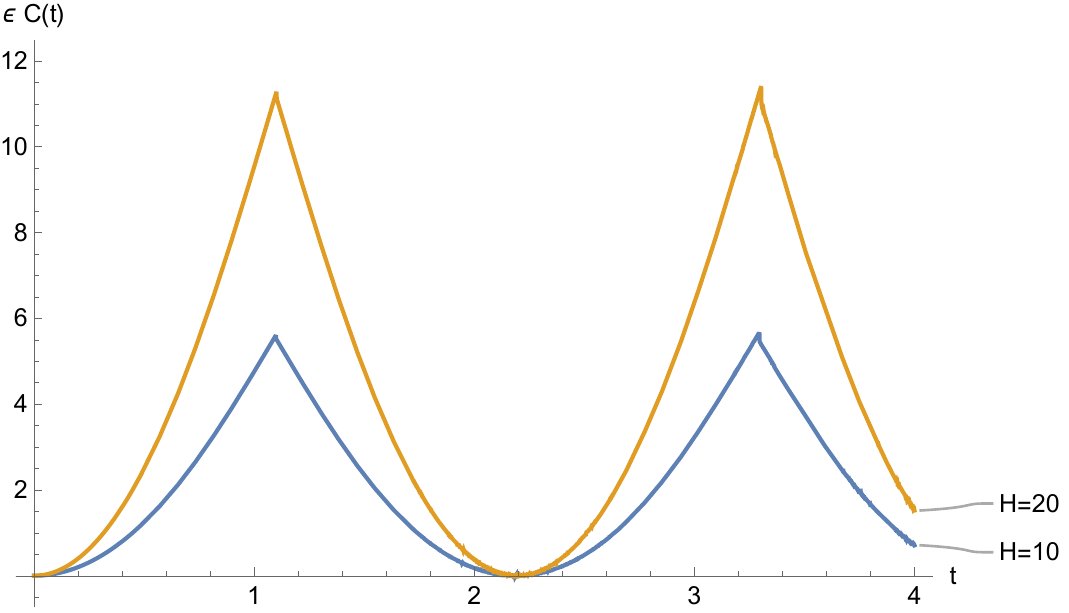}
    \caption{ $\epsilon \;\mathcal{C}(t)$, for $Q=1, \mu=1$ and $l=1$ and $\mathcal{H}=10,20.$}
    \label{fig:CtQmu}
\end{figure}

A plot for the complexity for $Q=1, \mu=1$ and $\mathcal{H}=10,20$  is provided in Figure \ref{fig:CtQmu}.

\bibliographystyle{JHEP}
\bibliography{main}

\end{document}